\documentclass[prx,aps,twocolumn,preprintnumbers,amsmath,amssymb,superscriptaddress,showpacs]{revtex4-2}

\usepackage{graphicx}
\usepackage{epstopdf}
\usepackage{amsmath}
\usepackage{multirow}
\usepackage{xcolor}
\usepackage{ulem}
\usepackage{float}
\usepackage{multirow}
\usepackage{array}

\newcommand{\RNum}[1]{\uppercase\expandafter{\romannumeral #1\relax}}
\usepackage[sort&compress]{natbib}

\begin{document}

\title {Normal and superconducting properties of La$_3$Ni$_2$O$_7$}

\author{Meng Wang}
\affiliation{Guangdong Provincial Key Laboratory of Magnetoelectric Physics and Devices, School of Physics, Sun Yat-Sen University, Guangzhou, Guangdong 510275, China}

\author{Hai-Hu Wen}
\affiliation{National Laboratory of Solid State Microstructures and Department of Physics, Collaborative Innovation Center of Advanced Microstructures, Nanjing University, Nanjing 210093, China}

\author{Tao Wu}
\affiliation{Hefei National Research Center for Physical Sciences at the Microscale, University of Science and Technology of China, Hefei, Anhui 230026, China}

\author{Dao-Xin Yao}
\affiliation{Guangdong Provincial Key Laboratory of Magnetoelectric Physics and Devices, School of Physics, Sun Yat-Sen University, Guangzhou, Guangdong 510275, China}

\author{Tao Xiang}
\affiliation{Institute of Physics, Chinese Academy of Sciences, Beijing 100190, China}
\affiliation{School of Physics, University of Chinese Academy of Sciences}

\date{\today}

\begin{abstract}
 This review provides a comprehensive overview of current research on the structural, electronic, and magnetic characteristics of the recently discovered high-temperature superconductor La$_3$Ni$_2$O$_7$ under high pressures. We present the experimental results for synthesizing and characterizing this material, derived from measurements of transport, thermodynamics, and various spectroscopic techniques, and discuss their physical implications. We also explore theoretical models proposed to describe the electronic structures and superconducting pairing symmetry in La$_3$Ni$_2$O$_7$, highlighting the intricate interplay between electronic correlations and magnetic interactions. Despite these advances, challenges remain in growing high-quality samples free of extrinsic phases and oxygen deficiencies and in developing reliable measurement tools for determining diamagnetism and other physical quantities under high pressures. Further investigations in these areas are essential to deepening our understanding of the physical properties of La$_3$Ni$_2$O$_7$ and unlocking its superconducting pairing mechanism.
\end{abstract}


\maketitle

\section{Introduction}

 Recently, superconductivity exceeding 80 K was found in La$_3$Ni$_2$O$_7$, which contains a bilayer of Ni-O planes~\cite{Sun2023}. This marks the discovery of the third high-$T_c$ superconductor associated with the transition metal elements, following copper-oxide and iron-based superconductors. So far, superconductivity in this material can only be achieved under certain pressures, typically above 10 GPa. Zero resistance was observed in selected single crystals \cite{Zhang2023, Hou2023, Zhou2023, Li2024} and polycrystalline samples \cite{Wang2023, Wang2024} under hydrostatic pressures.

 La$_3$Ni$_2$O$_7$ is a Ruddlesden-Popper material of layered nickelates, similar to cuprate superconductors. The structure of La$_3$Ni$_2$O$_7$ consists of alternating layers of LaO and NiO planes, stacked along the c-axis, promoting two-dimensional electronic behavior. At ambient pressure, La$_3$Ni$_2$O$_7$ adopts an orthorhombic structure. However, at high pressures above $\sim 10$ GPa, it undergoes a structural transition. This pressure-induced {\it Amam} to {\it I4/mmm} structural phase transition seems to be a prerequisite for the observation of superconductivity~\cite{Wang2024str, Wang2023, Li2024}.

 The superconducting volume fraction is a critical parameter that quantifies the extent of a material exhibiting superconductivity, typically measured by the fraction of diamagnetic susceptibility. Like many other superconductors in their early stage of discovery, the superconducting volume fraction of La$_3$Ni$_2$O$_{7}$ initially measured was low due to the poor quality of laboratory-synthesized samples, which often contain mixed phases, inhomogeneities, and randomly scattered oxygen defects. These imperfections cause inconsistencies in the reported superconducting volume fraction, with some studies observing fractions as low as 1\%~\cite{Zhou2023}, suggesting filamentary superconductivity, and some as high as 48\%~\cite{Li2024}, indicating bulk superconductivity. Recently,  Cheng and coworkers pointed out that this low volume fraction may result from a mixture of La$_4$Ni$_3$O$_{10}$ and La$_3$Ni$_2$O$_{7}$ phases. By replacing one-third of La with Pr, they significantly suppressed the trilayer phase, achieving a superconducting volume fraction of larger than 40\% in La$_2$PrNi$_2$O$_{7}$~\cite{Wang2023}.

 Although the observed superconducting volume fraction is low, the currently available samples have revealed some intriguing properties of this material. First, the onset superconducting transition temperature ($T_c$) is notably robust against an applied magnetic field~\cite{Sun2023,Zhang2023, Hou2023, Zhou2023, Li2024,Wang2023,Wang2024,Zhang2024, Sakakibara2024}, similar to the behavior of cuprates. This stability, unaffected by the low superconducting volume fraction, implies a short superconducting coherence length. Second, a linear temperature-dependent resistance has been observed above $T_c$ by several research groups ~\cite{Sun2023, Zhang2023}, even with samples of varying quality. This observation is puzzling because the resistance is contributed by both superconducting and non-superconducting phases, typically resulting in a more complex temperature behavior rather than a simple linear-T behavior. This puzzling observation could be understood in the following two cases: (1) both the superconducting and non-superconducting phases in the sample exhibit linear-T resistivity, and (2) if the superconducting phase exhibits linear-T behavior while the non-superconducting phase insulates. A linear-T dependence of resistivity is generally a signature of strong correlation~\cite{Yuan2022}, indicating an unconventional nature of superconductivity in La$_3$Ni$_2$O$_7$.

 La$_3$Ni$_2$O$_7$ exhibits complex spin and charge density wave behaviors that are significantly influenced by temperature and pressure~\cite{Liu2023,Zhang2023,wutao26}, which are crucial in understanding its electronic properties. Magnetic fluctuations are particularly critical for understanding the pairing mechanism of superconductivity. Various measurements, including Resonant Inelastic X-ray Scattering (RIXS)~\cite{wutao11}, muon-spin relaxation ($\mu$SR)\cite{RN790,wutao26}  and Nuclear Magnetic Resonance (NMR)~\cite{wutao16}, have revealed the presence of spin density waves (SDW) at $(\pi/2, \pi/2, \pi)$ in La$_3$Ni$_2$O$_7$ at ambient pressure and low temperatures. These SDWs indicate a strong interplay between magnetic and electronic properties. However, it remains unknown how SDW changes in the superconducting phase after the material undergoes a structural phase transition under high pressures.

 Theoretical research has been equally active, aiming to elucidate the superconducting mechanism in La$_3$Ni$_2$O$_7$~\cite{yaoprlmodel, dagotto23prb, wangqh2023,gu2023effective, Wu2024,yaohighTc, PhysRevLett.132.146002, PhysRevLett.132.036502, RN736, ouyang2023hund, PhysRevLett.131.236002, Jiang__2024,liu2023role, yang2023interlayer}. 
 First-principle density functional theory calculations suggest that at ambient pressure, two bands, whose low-energy excitations are primarily contributed by Ni $3d_{x^2-y^2}$ and in-plane O $2p$ orbitals,  cross the Fermi level~\cite{Sun2023}. The electronic band structure changes when pressure is applied, bringing a third band, mainly involving Ni $3d_{z^2}$ orbitals, to the Fermi surface~\cite{Sun2023,yaoprlmodel}. Some models attribute the pairing interaction to interlayer magnetic interactions between Ni $3d_{z^2}$ spins~ \cite{wangqh2023, gu2023effective, yaohighTc, PhysRevLett.131.236002, ShenQin2023, yang2023interlayer}. Other models, on the other hand, emphasize the role of electronic correlations and the significance of nickel-oxygen planes in facilitating superconducting pairing~\cite{yaohighTc, Jiang__2024, liu2023role, RN736, PhysRevLett.132.146002, PhysRevLett.132.036502, ouyang2023hund}. 
 
 This article provides a review of the current research on La$_3$Ni$_2$O$_7$ as a high-temperature superconductor under high pressures. We begin by discussing the material's synthesis and the structural phase transitions induced by pressure. Following this, we examine experimental results obtained using various techniques, including magnetic susceptibility, resistivity, Hall coefficient, and a range of spectral techniques, such as angular resolved photoemission spectroscopy (ARPES), NMR, $\mu$SR, RIXS, and optical conductivity. Finally, we explore the theoretical models proposed to explain the observed phenomena and the pairing symmetries predicted by different models. Through this review, we hope to provide a timely and critical perspective on La$_3$Ni$_2$O$_7$, highlighting its significance in the ongoing quest to discover new high-temperature superconductors.

\section{Material syntheses} 
\label{Sec:Mat}

\subsection{Polycrystalline samples}

 The Ruddlesden-Popper (RP) phases of the ternary La-Ni-O system can be described by the general formula  La$_{n+1}$Ni$_n$O$_{3n+1}$ with $n=1, 2, 3, ..., \infty$. The $n^{th}$ RP phase is characterized by stacked structures that interleave $n$ layers of perovskite-like LaNiO$_3$ with one layer of rocksalt-type LaO. The two end members of the RP phases, LaNiO$_3$ and La$_2$NiO$_4$, were first reported by A. Wold et al. in 1957 \cite{Wold1957} and A. Rabenau et al. in 1958 \cite{Rabenau1958}. 

 LaNiO$_3$ is a perovskite structure with the high oxidation state of Ni$^{3+}$ and can be synthesized by mixing La$_2$O$_3$ and NiO at 800$^\circ$C in a NaCO$_3$ flux. By replacing NaCO$_3$ with KOH, LaNiO$_3$ polycrystalline samples can be obtained at 400$^\circ$C, as molten KOH provides O$^{2-}$ ions, enabling dissolution of the metal oxides and stabilization of the higher oxidation states of Ni ions \cite{Shivakumara2003}. On the other hand, La$_2$NiO$_4$ with $n=1$ hosts the most stable oxidation state of Ni$^{2+}$ and can be readily synthesized by both solid-state reaction and optical floating zone (OFZ) methods \cite{Rabenau1958}. Intermediate RP phases of the La-Ni-O system were observed during synthesizing the $n=1$ and $n=\infty$ phases \cite{Wold1959}.  

 \begin{figure}[b]
  \includegraphics[width=0.9\columnwidth]{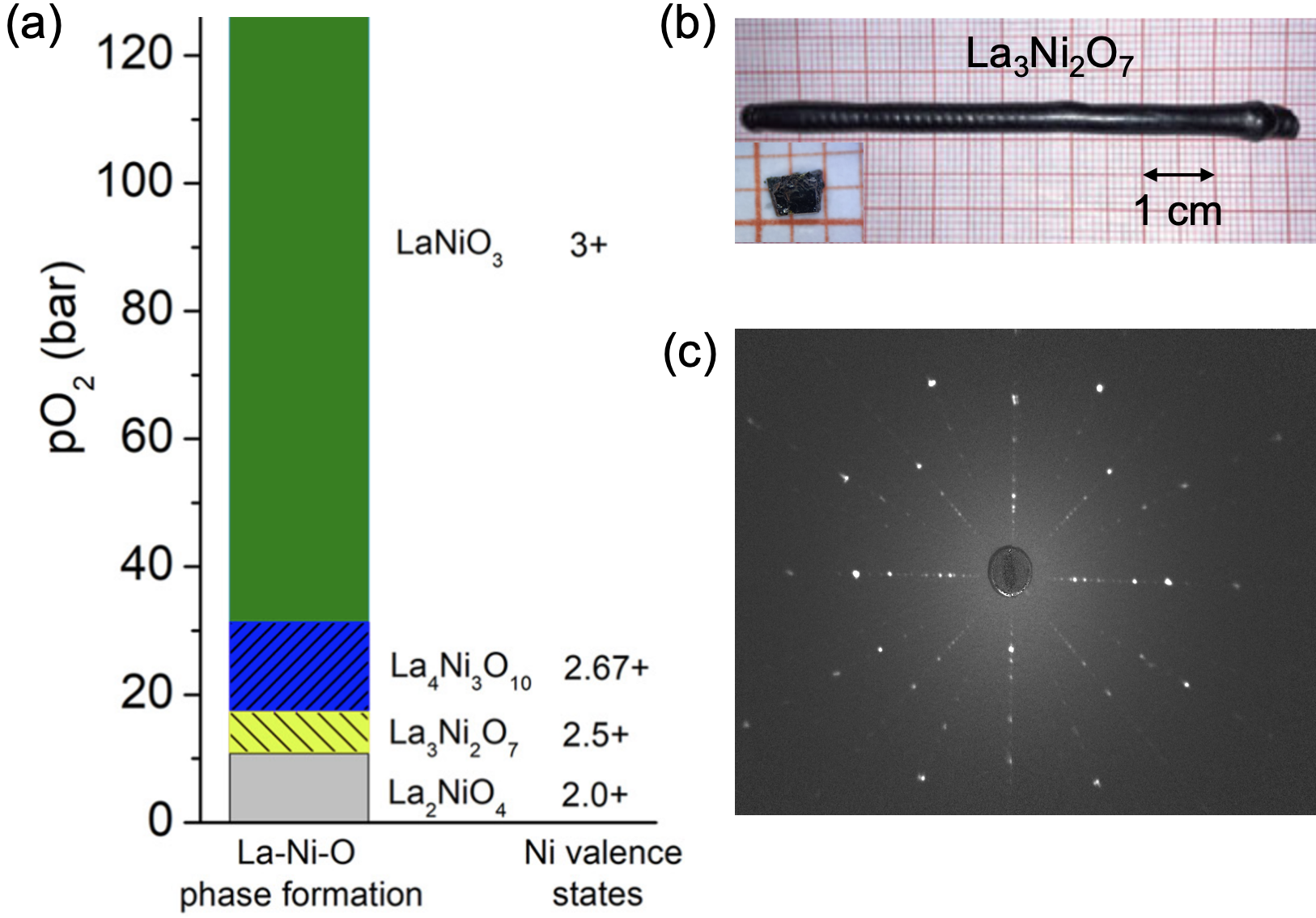}
    \caption{\label{fig:W1} (a) Relationship between the oxygen pressure and the formation of the La-Ni-O RP phases. Reprinted from Ref.~\cite{Zhang2020}. (b) A photo of the La$_3$Ni$_2$O$_7$ single crystal grown by the high-pressure optical floating zone furnace, reprinted from Ref.~\cite{Liu2023}. The inset image is a piece of cleaved single crystal. (c) A Laue pattern of the $ab$ plane of a single crystal.}
\end{figure}

 \begin{figure*}[t]
  \includegraphics[width=1.7\columnwidth]{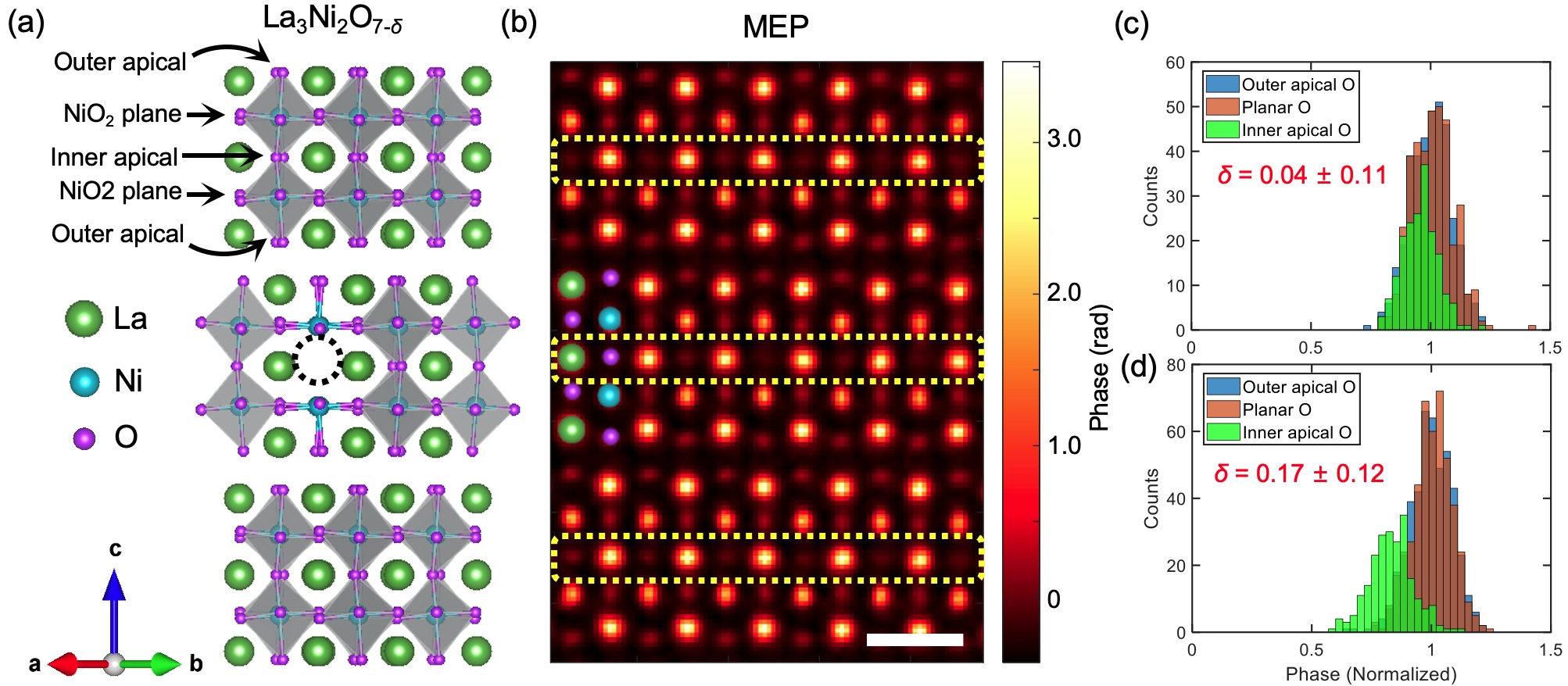}
    \caption{\label{fig:W2} (a) Crystal structure of La$_3$Ni$_2$O$_7$. The dashed circle represents an oxygen vacancy. (b) Projected multislice electron ptychography phase image from a region with an oxygen vacancy of $\sigma=0.4$. The dashed yellow rectangulars mark the inner apical oxygen sites with oxygen vacancies. (c) and (d) show the estimated oxygen vacancies for different sites of La$_3$Ni$_2$O$_{7-\delta}$. Reprinted from Ref.~\cite{Dong2023}. }
\end{figure*}
 The research on the RP phases with $n=2$ and 3 was initialized by C. Brisi et al. in 1981~\cite{Brisi1981}. Their experiment aimed to produce LaNiO$_3$ by heating a mixture of La(NO$_3$)$_3$ and Ni(NO$_3$)$_2$ powders in the air at 800$^\circ$C for 100 hours. However, further heating of the as-grown LaNiO$_3$ powders at 1100$^\circ$C for about 30 minutes led to the formation of a mixture of NiO and La$_4$Ni$_3$O$_{10}$, revealing that La$_4$Ni$_3$O$_{10}$ is thermodynamically stable in this condition. In contrast, when La(NO$_3$)$_3$ and Ni(NO$_3$)$_2$ were directly heated at 1100$^\circ$C for 24 hours with an atomic ratio of La$:$Ni of $4:3$, the bilayer RP phase of La$_3$Ni$_2$O$_7$ with a small amount of NiO was identified. It is worth mentioning that La$_3$Ni$_2$O$_7$ is a metastable phase at 1100$^\circ$C, as a gradual transformation of La$_3$Ni$_2$O$_7$ into La$_4$Ni$_3$O$_{10}$ can be observed through x-ray diffraction analysis. Further investigation using electron microscopy revealed the intergrowth of La$_2$NiO$_4$, La$_3$Ni$_2$O$_7$, and La$_4$Ni$_3$O$_{10}$ through the reaction of La$_2$O$_3$ and Ni(NO$_3$)$_2$ at a molar ratio of $3:4$ at a temperature of 1150$^\circ$C in the air for 5 hours \cite{Drennan1982}.

Decomposing La(NO$_3$)$_3$ and Ni(NO$_3$)$_2$ in nitric acid with a stoichiometric ratio, followed by heating the resultant compressed powders to  1150$^\circ$C can significantly improve the quality of La$_3$Ni$_2$O$_7$~\cite{Ram1986, Sreedhar1994, Zhang1994, Taniguchi1995}, allowing large areas of the single La$_3$Ni$_2$O$_7$ phase to be visualized by high-resolution electron microscopy (HREM) \cite{Ram1986}. Furthermore, a sol-gel method was employed by adding an equivalent molar proportion of citric acid to La$_2$O$_3$ and NiO after dissolving stoichiometric amounts of La$_2$O$_3$ and NiO in a nitric acid solution. This method was used to grow the polycrystalline samples of $n=2$ and $3$ RP members \cite{Carvalho1997, Wang2024}. Higher RP phases with $n=4$ and 5 were observed occasionally in the boundary of intergrowth phases or samples with small grain sizes \cite{Drennan1982, Ram1986}.

\subsection{Single crystals}

 The OFZ method has been instrumental in growing high-quality and large single crystals of copper oxide superconductors. This method has also been employed to grow single crystalline samples of the RP phases of nickelate. Due to the high valence states of nickel, the crystal growth process requires high-pressure oxygen gas \cite{Zhang2020}. 
 
 The standard solid-state reaction method can be used to synthesize the precursors for OFZ growth. Through multiple reactions of La$_2$O$_3$ and NiO at 1100$^\circ$C, a La$_3$Ni$_2$O$_{7-\delta}$ phase-dominated precursor with oxygen vacancies can be obtained. The  La$_3$Ni$_2$O$_{7-\delta}$ powders can be pressurized into rods with 10 cm in length and 5 mm in diameter. Annealing the rods at 1400$^{\circ}$C is necessary to improve the density of the feed and seed rods. This annealing process causes the precursor to deform into a mixture of La$_2$NiO$_4$, La$_3$Ni$_2$O$_{7-\delta}$, La$_4$Ni$_3$O$_{10-\delta}$, La$_2$O$_3$, and NiO. The oxygen pressure is crucial for the formation of specific RP phases of nickelates during the OFZ growth, as shown in Fig. \ref{fig:W1}. Successful growth of La$_3$Ni$_2$O$_{7-\delta}$ has been achieved with an oxygen pressure of $p$(O$_2$) = 15 bar \cite{Liu2023}. However, impurity phases such as La$_2$NiO$_4$ and La$_4$Ni$_3$O$_{10-\delta}$ may also be present using the same condition, indicating the influence of other parameters such as lamp power, movement speed, rotation speed, and precursor diameter. Under similar conditions, other groups obtained an alternating single-layer-trilayer stacking structure (1313) with the same chemical formula, La$_3$Ni$_2$O$_7$, as the bilayer structure (2222)~\cite{Puphal2023, Chen2024}. Additionally, some stacking faults were commonly observed by TEM in the RP phase structures.

 The cutting-edge multislice electron ptychography (MEP) TEM technique has been adopted to visualize oxygen microscopically in single crystals of the bilayer structural La$_3$Ni$_2$O$_{7-\delta}$~\cite{Dong2023}. The results reveal the inhomogeneity of oxygen vacancies in a length scale of 40 \AA. The oxygen vacancies $\delta$ spanning from 0 to 0.34 are mainly located on the inner-apical oxygen site, as shown in Fig. \ref{fig:W2}.

\section{Transport properties}

\begin{figure}[b]
  \includegraphics[width=0.9\columnwidth]{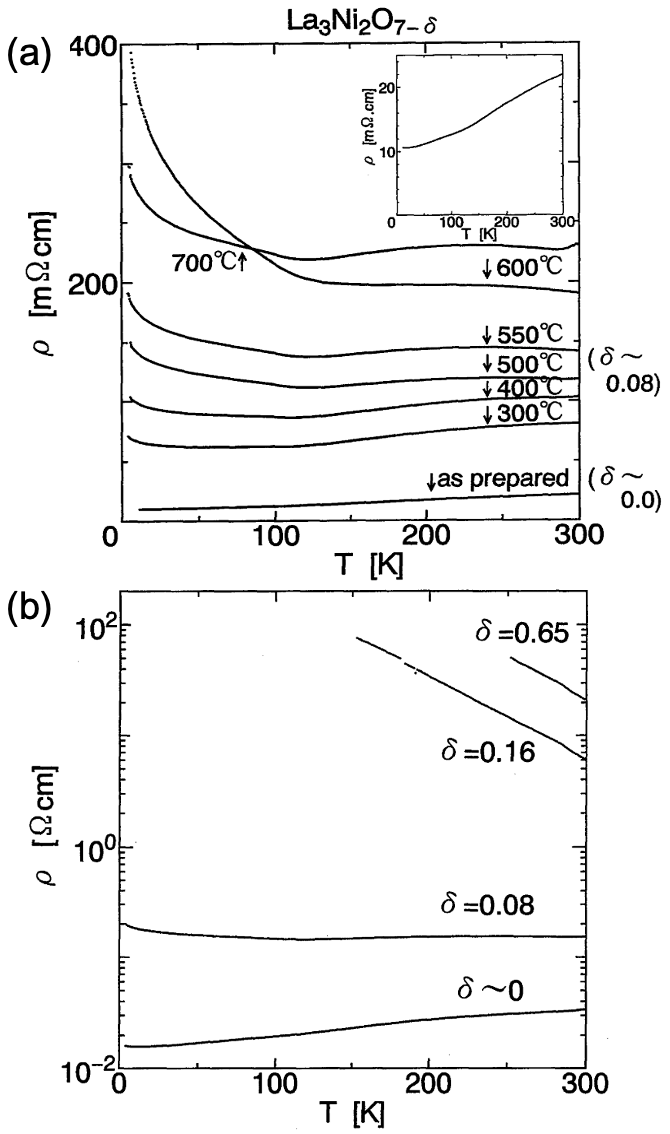}
    \caption{\label{fig:X1} Temperature dependence of resistivity of La$_3$Ni$_2$O$_{7-\delta}$ polycrystalline samples synthesized (a) at different temperatures in air and (b) with different oxygen content levels. Reprinted from Ref.~\cite{Taniguchi1995}.}
\end{figure}

Characterizations of La$_3$Ni$_2$O$_{7-\delta}$ in the early days were based on polycrystalline samples. The oxygen content of polycrystalline samples can be adjusted through post-annealing in different atmospheres, such as O$_2$, air, or H$_2$/Ar at approximately 450$^\circ$C. Thermogravimetric analysis (TGA) and powder neutron diffraction (PND) refinement are used to determine the average oxygen composition of the samples. Powder x-ray diffraction (XRD) measurements on the $\delta\approx0$ compounds suggested an orthorhombic structure (space group $Fmmm$) where the bonds of Ni$-$O$-$Ni along the $c$ axis are straight~\cite{Zhang1994}. Later on, PND revealed the structure is in the space group $Amam$, which is a nonstandard setting of $Cmam$ by allowing the normal direction of the NiO$_2$ plane as the $c$ axis. The angle of Ni$-$O$-$Ni along the $c$ axis is 168$^\circ$C for $Amam$, differing from the 180$^\circ$C angle of the $Fmmm$ space group~\cite{Ling1999}. A recent PND investigation indicates that both $Fmmm$ and  $Amam$ phases exist in the polycrystalline samples synthesized by the solid-state reaction~\cite{Xie2024}.

 By annealing the as-grown samples under the atmosphere of flowing Ar with 20$\%$ H$_2$ at 450$^\circ$C for 3 h~\cite{Taniguchi1995}, La$_3$Ni$_2$O$_{6.35}$ can be obtained. PND measurements did not observe any evidence of magnetic orders. It is worth noting that while different patterns of charge density waves (CDW) have been proposed for La$_3$Ni$_2$O$_{7-\delta}$ as a function of oxygen vacancies, no direct experimental observation of the CDW has been reported~\cite{Taniguchi1995}.

\begin{figure}[b]
  \includegraphics[width=0.8\columnwidth]{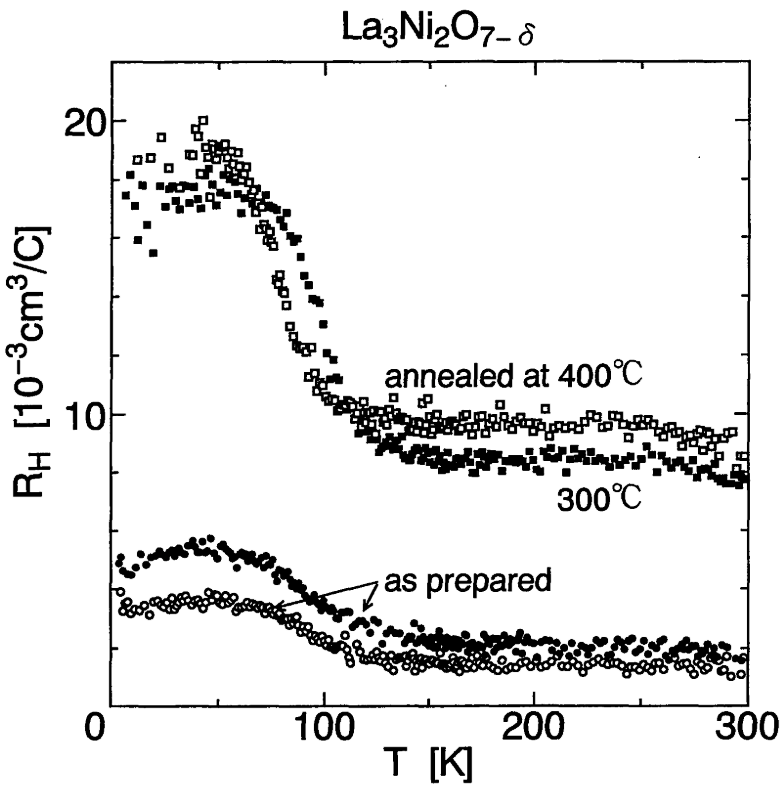}
    \caption{\label{fig:X2}  Temperature dependence of Hall coefficient for La$_3$Ni$_2$O$_{7-\delta}$ polycrystalline samples. Among the samples annealed at 300$^\circ$C-400$^\circ$C, the oxygen deficiency $\delta$ is below 0.08, and the Hall coefficient increases by about five times. Reprinted from Ref.~\cite{Taniguchi1995}.  }
\end{figure}

 Figure \ref{fig:X1} illustrates the temperature dependence of resistivity of La$_3$Ni$_2$O$_{7-\delta}$. It is evident that at $\delta=0$, the sample showed metallic behavior similar to that reported in Ref.~\cite{Liu2023} for single crystal samples at ambient pressure. Earlier studies found that polycrystalline samples, after heat treatment between 300$^\circ$C-500 $^\circ$C, experienced a rapid increase in resistivity, with a metal-insulator transition occurring around 120 K, accompanied by an upturn in resistivity at low temperatures, as shown in Fig. \ref{fig:X1}~\cite{Taniguchi1995}. This behavior intensified with further heat treatment at 600$^\circ$C-700$^\circ$C, with a resistivity increase of about three to four orders of magnitude compared to untreated samples. This increased resistivity was accompanied by a rapid enhancement in insulating behavior at low temperatures. High-pressure electric transport measurements were conducted on polycrystalline samples of La$_3$Ni$_2$O$_{7-\delta}$. The resistance shows weak insulating behavior up to 18.5 GPa. Oxygen vacancies are evident in samples with the upturn of resistance observed at ambient pressure~\cite{Hosoya2008}.

\begin{figure*}[t]
  \includegraphics[width=1.5\columnwidth]{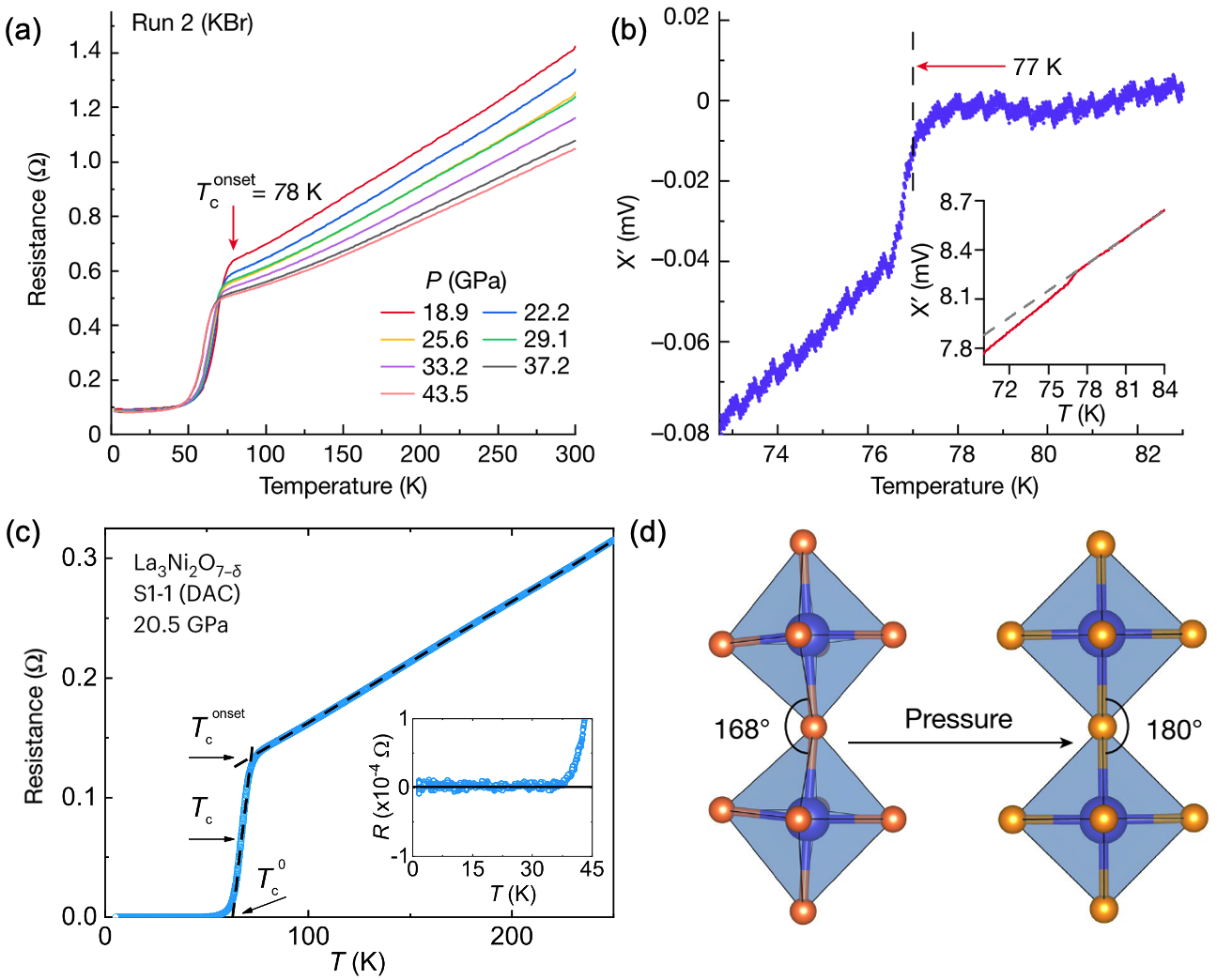}
    \caption{\label{fig:W3} (a) Resistance and (b) $ac$ magnetic susceptibility of La$_3$Ni$_2$O$_7$ under high pressure. (c) Measured zero resistance of La$_3$Ni$_2$O$_7$ at 20.5 GPa. (d) A sketch of the structural transition of the NiO$_6$ octahedra from ambient to high pressure. Reprinted from Refs.~\cite{Sun2023,Zhang2023}.}
\end{figure*}

 It is clear that the electronic behavior of this system undergoes rapid changes. In samples with higher oxygen content ($\delta \leq 0.08$), recent spectroscopic results indicated that charge and spin density waves occur at approximately 120 K and 150 K~\cite{RN785,RN790,wutao11,wutao16,wutao26}, respectively, consistent with the transport measurements on single crystal samples~\cite{Liu2023}. Figure~\ref{fig:X2} shows the temperature dependence of the Hall coefficient for samples with different oxygen contents. The positive Hall coefficient indicates hole-dominated charge carriers. Furthermore, there is a sudden increase in the Hall coefficient below approximately 120 K, suggesting a decrease in carrier concentration~\cite{Taniguchi1995,RN947}. 

Density Functional Theory (DFT) calculations~\cite{RN736,Pardo2011,Nakata2017} revealed that in the La$_3$Ni$_2$O$_7$ system, the bands near the Fermi level are primarily composed of the Ni-3$d_{x^2-y^2}$ and Ni-3$d_{z^2}$ orbitals with average 0.5 and 1 occupied electrons, respectively. The interlayer coupling splits the bonding and antibonding bands. Moreover, there are two Ni-3$d_{x^2-y^2}$-dominant Fermi surfaces at ambient pressure: an electron-like $\alpha$ Fermi surface around the $\Gamma$ point and a hole-like $\beta$ Fermi surface around the M point resembling the Fermi surface in cuprates. 

For as-prepared samples, the Hall coefficient at high temperatures is approximately 1$\sim$2$\times$10$^{-3}$cm$^{3}$/C, with hole carriers around 3$\sim$6$\times$10$^{20}/$cm$^{3}$, roughly 0.25$\sim$0.5 holes/Ni. If we assume that the d$_{z^2}$ orbital is half-filled by one electron and thus localized due to the strong correlation effect, we are left with 0.5+$\delta$ electrons in the d$_{x^2-y^2}$ orbital. Thus, in the lower Hubbard band of d$_{x^2-y^2}$ orbital, we have 0.5-$\delta$ holes, which is close to the experimentally observed value. Therefore, Hall effect measurements indicate that the electronic behavior of the d$_{z^2}$ orbitals may be localized, and the conductivity is primarily contributed by the d$_{x^2-y^2}$ orbital, roughly consistent with the ARPES experiment at ambient pressure~\cite{wutao1}. When the temperature is lowered for the as-prepared sample, the Hall coefficient $R_H$ gets enhanced below about 120 K, showing the signature of a density wave gap near Fermi energy. As the oxygen content decreases to $\delta\ge0.08$, the Hall coefficient increases about five times, which is not accountable by a simple counting of the nominal hole number (0.5-$\delta$)/Ni. Combining this fact and the rapidly enhanced resistivity as well as the strong insulating behavior in an oxygen-deficient sample ($\delta\ge0.08$), it is reasonable to conclude that the conduction of electrons in the d$_{x^2-y^2}$ orbital is strongly disturbed through Hund's coupling with the d$_{z^2}$ orbital.

\section{Superconducting transition}

 Sun et al. reported a signature of superconductivity with a transition temperature around 80 K in single crystals of La$_3$Ni$_2$O$_7$ above 14 GPa in 2023~\cite{Sun2023}. Both drops in resistance and alternating current ($ac$) magnetic susceptibility were observed [Fig.~\ref{fig:W3}(a) -(b)]. A structural transition from the low-pressure $Amam$ phase to the high-pressure $Fmmm$ phase was suggested~\cite{Wang2024str}. Based on the DFT calculations, only the bands dominated by the 3$d_{x^2-y^2}$ orbitals appear on the Fermi surface at ambient pressure. The 3$d_{z^2}$ $\sigma$-bonding band emerges above the Fermi level after the $Amam$-$Fmmm$ structural transition by pressure.  The structural transition flattens the Ni-O-Ni bonding angle along the $c$ axis as shown in Fig.~\ref{fig:W3}(d), which enhances the interlayer antiferromagnetic exchange interactions. Further investigations suggest that the high-pressure superconducting phase is in the tetragonal $I4/mmm$ space group~\cite{Wang2024str, Li2024, Wang2023}. 

 Zero resistance was observed in both single crystals [Fig. \ref{fig:W3}(c)] and polycrystalline samples of La$_3$Ni$_2$O$_7$ after the report of superconductivity \cite{Zhang2023, Hou2023, Zhou2023, Wang2023, Wang2024, Li2024}. The $ac$ diamagnetic susceptibility, reflecting the Meissner effect of superconductivity, was also confirmed on single crystal samples \cite{Sun2023, Zhou2023, Li2024}. However, the superconducting volume fraction is relatively low and varies with samples, indicating that samples are highly inhomogeneous~\cite{Zhou2023}. A homogeneous hydrostatic pressure is also crucial for the observation of zero resistance. In some samples, only a weak resistance drop was observed around 80 K under high pressure~\cite{Zhang2024, Sakakibara2024}. TEM measurements on single crystals revealed the existence of different oxygen vacancies in different areas of samples, as shown in Fig. \ref{fig:W2}, especially for the inner apical oxygen sites that connect two NiO$_6$ octahedra along the $c$ axis~\cite{Dong2023}. The inner apical oxygen vacancies suppress the direct interlayer couplings~\cite{Xie2024}. 

 Based on a recent report, the pressure-driven superconductivity shows a phase region with a right-triangle shape and persists up to 90 GPa. The maximum superconducting volume fraction reported is $\sim$48$\%$, suggesting that the superconductivity originates from the bilayer La$_3$Ni$_2$O$_7$ phase~\cite{Li2024}. For comparison, the superconducting volume fraction is close to 100$\%$ in La$_4$Ni$_3$O$_{10}$~\cite{Zhu2023}. The fault stacking of the perovskite LaNiO$_3$ layers is unavoidable in synthesizing La$_3$Ni$_2$O$_7$. This extrinsic phase will inevitably suppress the superconducting volume fraction in La$_3$Ni$_2$O$_7$. La in La$_3$Ni$_2$O$_7$ can be partially replaced successfully by the isovalent Pr~\cite{Wang2023}. Nuclear quadrupole resonance (NQR) and TEM measurements reveal that Pr doping on the La sites can suppress the fault stackings effectively and result in a relatively purer bilayer phase of La$_{3-x}$Pr$_x$Ni$_2$O$_{7-\delta}$. The superconducting volume fraction is above $\sim$40$\%$ in La$_{2}$PrNi$_2$O$_{7-\delta}$, demonstrating a bulk superconductivity~\cite{Wang2023}.

\begin{figure}[t]
  \includegraphics[width=0.8\columnwidth]{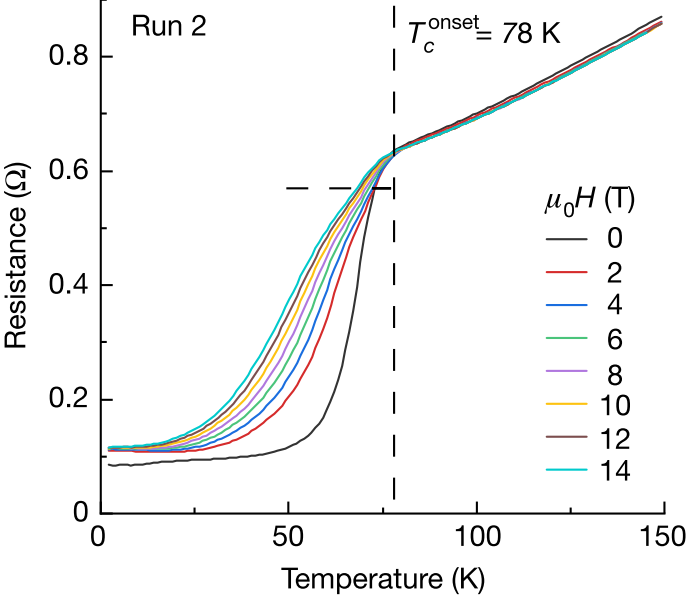}
    \caption{\label{fig:X.3} Variation in the resistance of La$_3$Ni$_2$O$_{7-\delta}$ across different magnetic fields. The horizontal dashed line indicates a 10\% reduction in resistance from the normal state, serving as the threshold to determine the upper critical field. Reprinted from Ref.~\cite{Sun2023}.}
    \end{figure}

 Although the superconducting volume fraction of La$_3$Ni$_2$O$_{7-\delta}$ is below 100\%, several groups have confirmed that its superconducting transition temperature exceeds 80 K \textcolor{red}~\cite{Zhang2024,Li2024}. Additionally, the resistance transition curve under an applied magnetic field behaves similarly to that of cuprates. The onset critical temperature, defined as the temperature at which resistance decreases to 90\% of its normal state value, diminishes gradually with increasing magnetic field. Figure~\ref{fig:X.3}  shows how the resistance varies with temperature in different applied fields at 18.9 GPa~\cite{Sun2023}.

 Several groups have investigated the resistance of materials under magnetic fields to determine how the upper critical field varies with temperature. They used the Ginzburg-Landau formula to fit the data and obtained relatively high values for the upper critical field at zero temperature. Specifically, the reported upper critical field, $\mu_0H_{c2}(0)$, are 186 T at 18.9 GPa with $T_c\sim$ 78 K in Ref.~\cite{Sun2023}, 103.3 T at 15 GPa with $T_c\sim$ 78.2 K in Ref.~\cite{Wang2023},  and 103.3 T at 20.5 GPa with $T_c\sim$ 66 K in Ref.~\cite{Zhang2023}. These upper critical field values could increase with further refinement in sample quality.

 The coherence length is related to the upper critical field by the formula $\xi(0)=\sqrt{\Phi/2\pi\mu_0H_{c2}(0)}$ at zero temperature in the Ginzburg-Landau theory, where $\Phi=h/2e$ represents the flux quantum. Using data reported in Ref.~\cite{Sun2023}, the in-plane coherence length is estimated to be 1.44 nm, which is relatively short compared to conventional phonon-mediated superconductors.

\begin{figure*}[htp]
    \centering
    \includegraphics[width=0.8\textwidth]{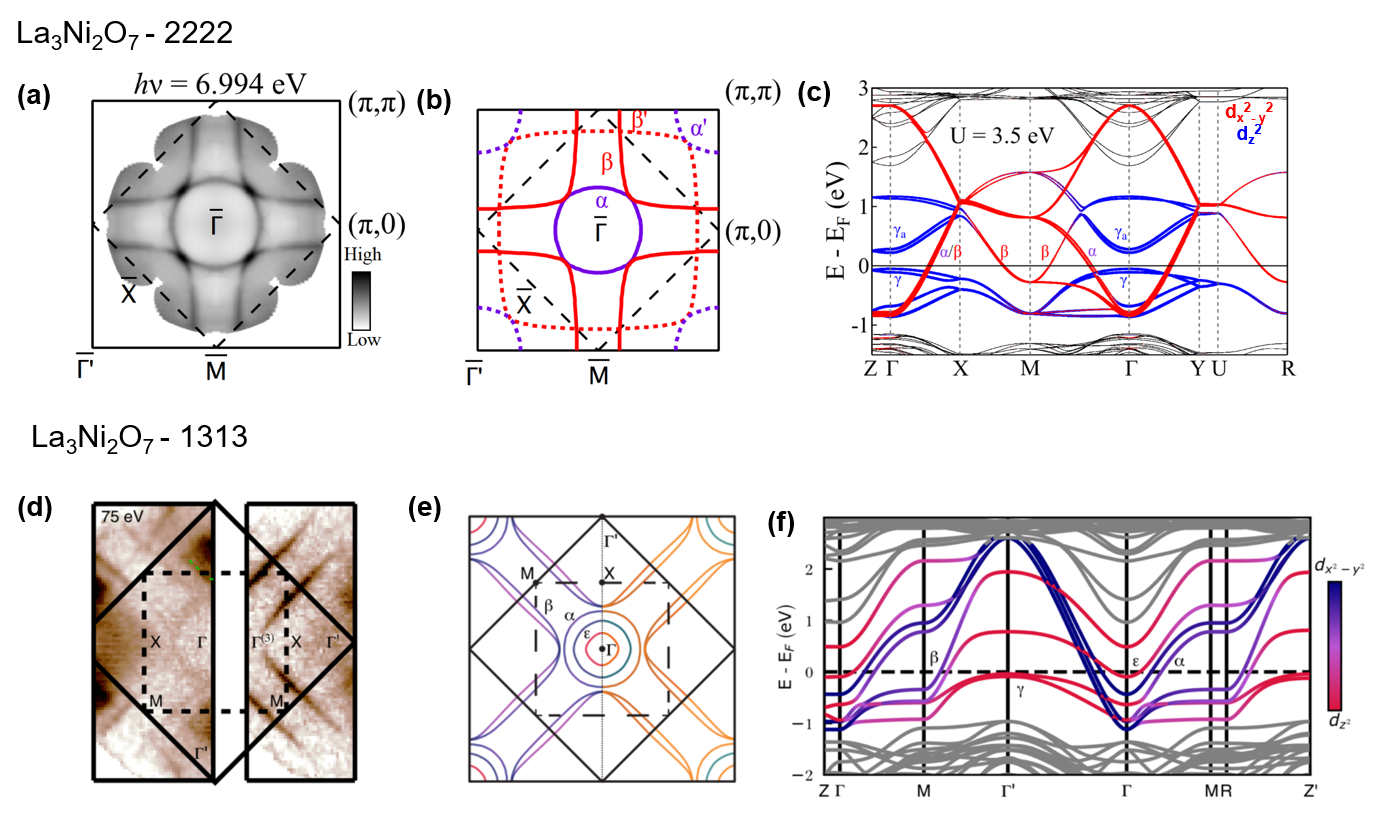}
    \caption{\label{fig: w1} (a-b) Fermi surfaces measured by ARPES and (c) band structures calculated by DFT with U = $3.5$ eV for La$_3$Ni$_2$O$_7$ (2222). Two Fermi surface sheets ($\alpha$ and $\beta$) are observed.  (d-e) Fermi surfaces and (f) band structure of La$_3$Ni$_2$O$_7$ (1313). \\
    (a-c) reprinted from \cite{wutao1}, (d-f) from \cite{wutao2}}
    \label{fig:twocolumn}
\end{figure*}

\section{Electronic structures}

 The electronic band structures of La$_3$Ni$_2$O$_7$ in two distinct structures, 2222 and 1313, have been investigated using ARPES~\cite{wutao1,wutao2}. Figure \ref{fig: w1} displays these measurements alongside results from DFT calculations. In both structures, three primary bands are identified near the Fermi level: the Ni-3$d_{x^2-y^2}$-derived $\alpha$ and $\beta$ bands, and the Ni-3$d_{z^2}$-derived $\gamma$ band, consistent with DFT predictions. The $\alpha$ and $\beta$ bands intersect the Fermi level, forming an electron-like sheet around $(0,0)$ and a hole-like sheet around $(\pi,\pi)$, respectively. Conversely, the $\gamma$ band is relatively flat around $(\pi,\pi)$ and lies approximately 50 meV below the Fermi level in the 2222 structure and 25 meV in the 1313 structure. This Fermi surface topology bears similarities to that of the trilayer nickelate La$_4$Ni$_3$O$_{10}$, which also exhibits superconductivity under pressure~\cite{Sakakibara2024,wutao4,wutao5,Zhu2023}. In La$_4$Ni$_3$O$_{10}$, the $\gamma$ band intersects the Fermi level above the density-wave transition temperature, $T_{DW} \sim 135$ K. Below $T_{DW}$, the $\gamma$ band shifts to about 20 meV below the Fermi level. This shift is attributed to a gap opening resulting from the density-wave transition~\cite{wutao7}.

 The existing ARPES measurements for La$_3$Ni$_2$O$_7$ were conducted at low temperatures. Additional temperature-dependent ARPES studies are required to elucidate the evolution of the $\gamma$ band across the density-wave transition temperature. Understanding the evolution of the $\gamma$ band with pressure is also crucial for unraveling the mechanisms behind high-temperature superconductivity in this material~\cite{Sun2023}.

ARPES has also revealed substantial band renormalization effects that vary between the 2222 and 1313 structures (see Table~\ref{Table 1}). In the 2222 structure, the $\alpha$ and $\beta$ bands display relatively mild band renormalizations, approximately 2, which are nearly isotropic in momentum space, similar to those observed in La$_4$Ni$_3$O$_{10}$~\cite{wutao1,wutao7}. In contrast, the $\gamma$ band in the same structure undergoes stronger band renormalization, ranging from 5 to 8, which is also momentum-dependent. Conversely, in the 1313 structure, all bands experience strong and uniformly orbital band renormalization, ranging from 5 to 6.

\begin{table}[b]
\centering
\caption{Band renormalization factors determined by ARPES for 2222, 1313, and 4310 at ambient pressure.}
\label{Table 1}
\begin{tabular}{c|c|c}
\hline
Sample                         & Orbits             & Renormalization   factors \\
\hline
\multirow{2}{*}{2222} & d$_{x^2-y^2}$ ($\alpha$,$\beta$ band) & $\sim$2                   \\
                               & d$_{z^2}$ ($\gamma$ band)       & 5 $\sim$8     
                               \\
\hline
\multirow{2}{*}{1313} & d$_{x^2-y^2}$($\alpha$,$\beta$ band)  & 5 $\sim$5.5               \\
                               & d$_{z^2}$($\gamma$ band)        & 5 $\sim$6                 \\
\hline
4310                      & d$_{x^2-y^2}$ ($\alpha$,$\beta$ band) & 2 $\sim$2.5 \\   
\hline
\end{tabular}
\end{table}

\begin{figure}[b]
  \includegraphics[width=\columnwidth]{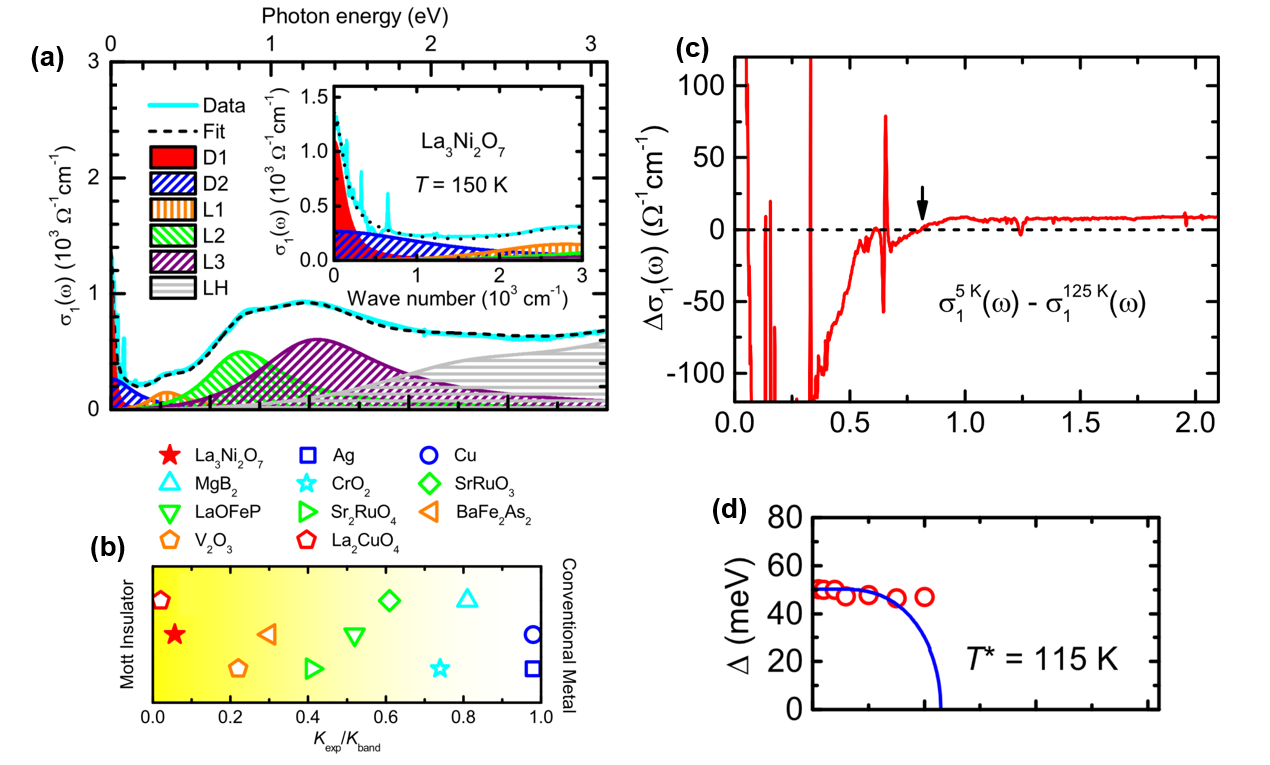}
    \caption{\label{fig: w2} (a) Optical conductivity  (cyan solid curve) and the Drude-Lorentz fitting result (black dashed line) of La$_3$Ni$_2$O$_7$. (b) $K_{exp}/K_{band}$ for La$_3$Ni$_2$O$_7$ (solid star) and some other materials. (c) Difference in the optical conductivity $\Delta \sigma_1(\omega)$ at $5$ K and $125$ K. The arrow indicates the zero-crossing point in $\Delta \sigma_1(\omega)$, which corresponds to the gap energy 2$\Delta$. (d) Temperature dependence of $\Delta$ (red open circles). The blue solid line denotes the mean-field behavior. Reprinted from \cite{RN785}. }
\end{figure}

In addition to ARPES experiments, optical conductivity measurements on La$_3$Ni$_2$O$_7$ have identified two Drude components, originating from bands primarily influenced by the Ni-3$d_{x^2-y^2}$ and Ni-3$d_{z^2}$ orbitals at the Fermi level~\cite{RN785}. The experimentally derived electron kinetic energy, $K_\text{exp}$, obtained by integrating the optical conductivity, is notably lower than the DFT result, $K_\text{DFT}$, (see Fig.~\ref{fig: w2}). Their ratio, $K_\text{exp}/K_\text{DFT}$, is approximately 0.072 for La$_3$Ni$_2$O$_7$, similar to that observed in doped cuprates.

\begin{figure*}[t]
  \includegraphics[width=1.7 \columnwidth]{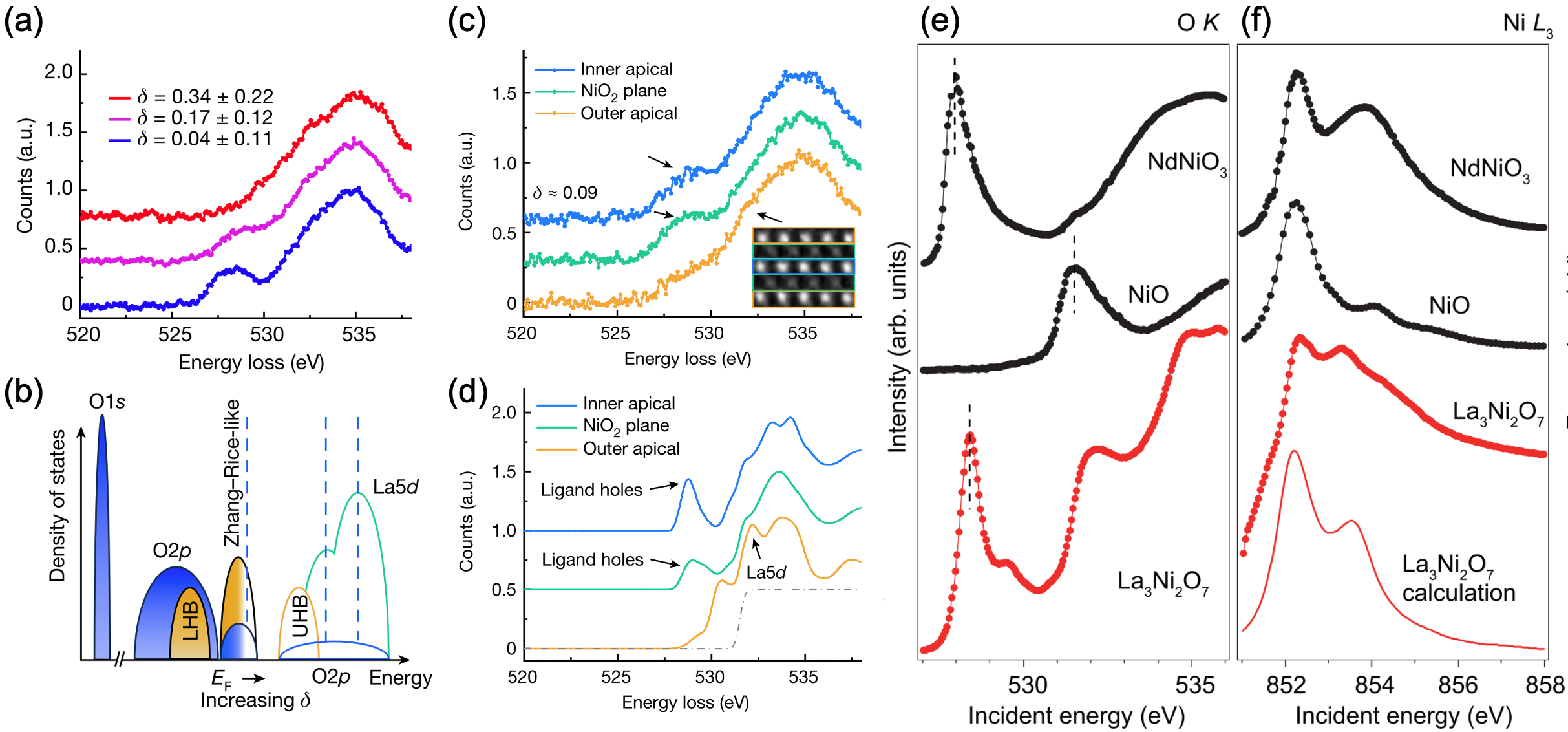}
    \caption{\label{fig:w3} (a) EELS results of La$_3$Ni$_2$O$_{7-\delta}$. Different color curves are from different regions of oxygen vacancies. (b) Illustration of the electronic structure of La$_3$Ni$_2$O$_{7-\delta}$ derived from EELS. (c) EELS on different oxygen sites of La$_3$Ni$_2$O$_{7-\delta}$ with $\delta=0.09$. (d) DFT calculated EELS spectra for the three distinct oxygen sites.  (e-f) XAS spectra of La$_3$Ni$_2$O$_7$ (red filled circles), NiO and NdNiO$_{3}$ (black filled circles), taken at the O $K$-edge and Ni $L_{3}$-edge, respectively. (a-d) are reprinted from \cite{Dong2023} and (e-f) from \cite{wutao11}. }
\end{figure*}

Electron energy loss spectroscopy (EELS) of the O-$K$ edge~\cite{Dong2023}, combined with X-ray absorption spectroscopy at the O $K$-edge and Ni $L_3$-edge~\cite{wutao11}, indicates that the O-$2p$ band in La$_3$Ni$_2$O$_7$ is partially filled, akin to cuprates within the Zaanen-Sawatzky-Allen framework~\cite{wutao37}. This contrasts with the electronic structure of infinite-layer nickelates~\cite{wutao38,wutao39}. Furthermore, high-resolution EELS at the atomic plane level ~\cite{Dong2023} revealed substantial $p-d$ hybridization in the $2p$ orbitals of both inner apical and planar oxygen atoms (Fig.~\ref{fig:w3}). In contrast, the outer apical oxygen is less relevant to low-energy physics and could be neglected in minimal theoretical analyses.

\begin{figure*}[t]
    \centering
    \includegraphics[width=1.7\columnwidth]{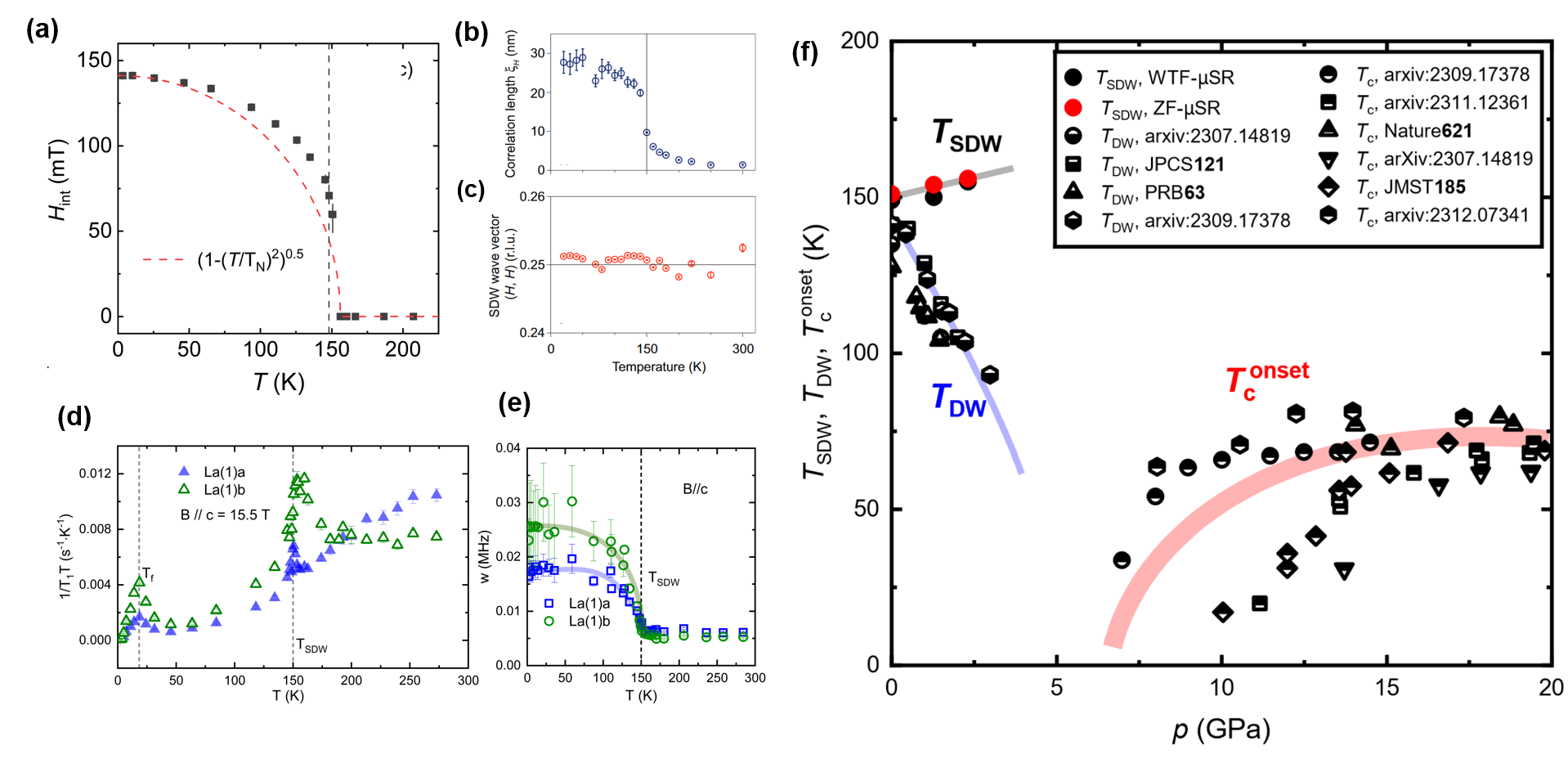}
    \caption{\label{fig:w4} (a) Temperature dependence of magnetic order of La$_3$Ni$_2$O$_7$, compared with the temperature dependence of $\sqrt{1-(T/T_N)^2}$. Temperature dependence of (b) the SDW correlation length and (c) the characteristic SDW wave vector. (d-e) Temperature dependence of $/T_1T$ and linewidth for La(1)a and La(1)b sites in an applied magnetic field parallel to the $c$-axis. (f) Pressure-temperature (p-T) phase diagram of polycrystalline La$_3$Ni$_2$O$_{7-\delta}$.  (a) is reprinted from \cite{RN790}, (b-c) from \cite{wutao11}, (d-e) from \cite{wutao16}, and (f) from \cite{wutao26}.
    }
\end{figure*}

\section{Density wave instabilities}

 The nature of the density-wave transition in La$_3$Ni$_2$O$_7$ at ambient pressure has attracted much interest in recent spectral studies. Theoretical calculations indicate that La$_3$Ni$_2$O$_7$ can exhibit SDW and CDW states under low~\cite{Seo1996} and high pressures, resulting from the scattering between the multiple Fermi surface sheets of La$_3$Ni$_2$O$_7$. Particularly, random phase approximation (RPA) calculations have identified specific regimes and predicted an SDW at the wave vector ($\pm\pi$/2, $\pm\pi$/2) ~\cite{yaoprlmodel} and a CDW at ($\pm\pi$, $\pm\delta$) with a small $\delta$~\cite{2023Christiansson}. These density-wave instabilities suggest the formation of diagonal stripes~\cite{PhysRevLett.131.236002}.
 
 Initial $^{139}$La NMR experiments conducted on polycrystalline samples of La$_3$Ni$_2$O$_7$ indicated a potential density-wave-like transition occurring below 150 K~\cite{wutao14, wutao12,wutao13}. However, whether it is an SDW or CDW remains to be discovered. 
 
 Recent experiments, including $\mu$SR, NMR, and RIXS measurements, have identified an SDW transition in La$_3$Ni$_2$O$7$ with a transition temperature $T_\text{SDW}$ around 150 K [Fig.~\ref{fig:w4} (a-e)].

 On the other hand, optical conductivity measurements on La$_3$Ni$_2$O$_7$ revealed an energy gap opened below 115 K, suggesting the formation of a CDW order in La$_3$Ni$_2$O$_7$  (Fig. \ref{fig: w2}). This gap opening temperature is significantly lower than the SDW transition temperature, unlike in La$_4$Ni$_3$O$_{10}$. ARPES measurements around this gap opening temperature could be used to elucidate the mechanisms driving this energy gap in La$_3$Ni$_2$O$_7$~\cite{wutao1, wutao2}.

 In the trilayer nickelate La$_4$Ni$_3$O$_{10}$, the density-wave transition results in an intertwined state exhibiting both charge and spin order below T$_{DW}$~\cite{wutao18}. This difference between La$_4$Ni$_3$O$_{10}$ and La$_3$Ni$_2$O$_7$ was confirmed by an ultrafast reflectivity measurement~\cite{wutao20}. The relaxation time in La$_4$Ni$_3$O$_{10}$ diverges near its density-wave transition temperature, contrasting sharply with the kink-like changes observed in La$_3$Ni$_2$O$_7$. Such divergence in relaxation times is a common feature observed in density-wave materials~\cite{wutao21, wutao22,wutao23, wutao24}, which can be understood using the Rothwarf-Taylor model~\cite{rt1967}. These findings suggest that while La$_4$Ni$_3$O$_{10}$ displays typical density-wave dynamics, La$_3$Ni$_2$O$_7$ exhibits a crossover behavior. 

 As no long-range magnetic order was observed in the neutron scattering measured down to 10 K, the SDW order is likely short-ranged or involving only small magnetic moments in La$_3$Ni$_2$O$_7$~\cite{Ling1999,Xie2024}. Intriguingly, a $\mu$SR experiment under pressure revealed a slight increase in the SDW transition temperature with pressure up to 2.3 GPa, as shown in Fig.~\ref{fig:w4}(f)~\cite{wutao26}, which contrasts with earlier transport measurements, indicating that density-wave transitions are typically suppressed with increasing pressure~\cite{Sun2023, Zhang2023, Wang2024, Zhou2023, Li2024,wutao32}. Further ultrafast optical spectroscopy measurements indicate that the density-wave-like transition observed at 150 K remains stable under pressure, persisting up to the maximum applied pressure of 34.2 GPa~\cite{Meng2024}.

  These diverging results strongly suggest the existence of two different density-wave orders in  La$_3$Ni$_2$O$_7$. With increasing pressure, the SDW order and CDW order evolve differently in pressure [Fig. \ref{fig:w4}(f)]. For comparison, spin and charge order are strongly intertwined and exhibit similar pressure dependence in high-T$_c$ cuprates. The peculiar pressure-dependence of density waves in La$_3$Ni$_2$O$_7$ is unseen before and deserves further investigation.

 While the existence of CDW order in La$_3$Ni$_2$O$_7$ remains debated, an SDW-type magnetic ground state has been directly observed by RIXS and other experimental measurements~\cite{RN790, wutao16, wutao11,wutao26, Xie2024}. Below the transition temperature $T_\text{SDW}$, Chen et al. reported that dispersive magnetic excitations peak at approximately 70 meV at the wave vectors $(0,0)$ and $(0.5,0)$, diminishing to zero energy at $(0.25,0.25)$~\cite{wutao11}. These observations strongly indicate the presence of quasistatic SDW ordering. Theoretical analysis suggests that the interlayer magnetic coupling is significantly larger than the intralayer ones, consistent with the recent inelastic neutron scattering measurement results~\cite{Xie2024}. The spectral weight center of the spin excitations, as measured by RIXS on single crystal samples, is 52 meV. In contrast, the center measured by inelastic neutron scattering on powder samples is 45 meV. This reduction in the spectral weight center can be attributed to the increased oxygen vacancies at the inner apical sites in the powder samples, which weaken the magnetic exchange interactions. These findings are essential for exploring the difference between La$_3$Ni$_2$O$_7$ and infinite-layer nickelates.

 Furthermore, several spin configurations corresponding to a spin order of $Q = (0.25,0.25)$ have been proposed~\cite{wutao11}. Three candidate stripe-type spin configurations have been explored through various spectral experiments~\cite{wutao11, wutao16}. Two of these configurations exhibit spin-charge stripe orders, with one of them closely resembling the pattern observed in La$_{1.5}$Sr$_{0.5}$NiO$_4$~\cite{wutao33}. The third configuration displays an SDW order without charge order. Despite additional insights from $\mu$SR and NMR experiments, which have narrowed down the potential spin configurations, the precise spin structure in La$_3$Ni$_2$O$_7$ remains debating~\cite{wutao16, wutao26}.

 Notably, a recent NMR study on single crystals of La$_3$Ni$_2$O$_7$ indicated significant sample inhomogeneity, leading to a remarkable splitting of $^{139}$La NMR spectra at the La(1) site within each bilayer~\cite{wutao16}. This inhomogeneity, attributed to an uneven distribution of inner apical oxygen~\cite{wutao16}, may profoundly influence the magnetic ordering moments. Further investigations using more homogeneous single crystals are imperative for accurately determining the spin configuration.

\section{Theoretical analyses }
\label{Sec:Theory}

\subsection{Theoretical models}

 Theoretical and numerical studies have played an important role in understanding the pairing mechanism in La$_3$Ni$_2$O$_7$. DFT calculations show that the low-energy bands contributed mainly by the two $e_g$ Ni $3d_{x^2-y^2}$ and $3d_{z^2}$ orbitals cross the Fermi level and the other bands from the $t_{2g}$ orbitals of Ni ions are far below the Fermi level in both high pressure $Fmmm$ and $I4/mmm$ phases~\cite{yaoprlmodel}. The superconductivity appears when the $\gamma$-band, which is mainly contributed by the $3d_{z^2}$ orbitals, emerges above the Fermi level, and the apical Ni-O-Ni bond approaches 180$^{\circ}$ angle in the high-pressure $Fmmm$ phase~\cite{Sun2023}. This $\gamma$ band is flat along the $\Gamma$-X and $\Gamma$-Y directions.

 In addition to the $\gamma$ band, the $\alpha$ and $\beta$ bands also intersect with the Fermi level~\cite{PhysRevB.109.L081105, dagotto23prb,PhysRevB.108.L201121,2023Christiansson,gu2023effective,RN736}. In these two bands, electronic states around the Fermi level contribute mainly by Ni $d_{x^2-y^2}$ and in-plane O $2p$ orbitals. Unlike the $\gamma$-band, these two bands cross the Fermi surfaces even in the low-pressure orthorhombic $Amam$ phase~\cite{wang2024electronic}. These calculated results are consistent with the ARPES measurement results~\cite{wutao1}. To quantitatively account for the observed Ni-$3d_{z^2}$ flat band about 50meV below the Fermi level, an effective on-site Coulomb interaction U (3 - 6eV) is needed in the LDA+U calculations, suggesting the importance of electron-electron correlations in La$_3$Ni$_2$O$_7$.

 In La$_3$Ni$_2$O$_7$, the oxygen ionization requires 14 electrons: the three La atoms contribute nine electrons from their valence shells, while the two Ni atoms supply the remaining five electrons. Each Ni atom is initially at the $3d^84s^2$ configuration, typically two Ni atoms donating four $4s$ electrons. Consequently, an average of 0.5 electrons per Ni atom must be removed from the $3d$ orbitals, resulting in a $3d^{7.5}$ configuration. For this $3d^{7.5}$ configuration, six electrons should fully occupy the three $t_{2g}$ orbitals, leaving average 1.5 electrons for the two $e_g$ ($3d_{x^2-y^2}$ and $3d_{z^2}$) orbitals. As the bilayer coupling will further split each $e_g$ orbital into two orbitals, the $3d_{x^2-y^2}$ and $3d_{z^2}$ orbitals will become on average 1/4-filled and half-filled, respectively. Incorporating Hund's coupling and electron correlations would further complicate this picture~\cite{PhysRevLett.132.146002, PhysRevLett.132.036502, PhysRevB.109.L081105, ouyang2023hund}. Hence, unlike cuprate superconductors, La$_3$Ni$_2$O$_7$ may display a multi-orbital behavior akin to iron-based superconductors. The interplay between these orbitals may contribute to various dynamic electronic phenomena such as spin density waves, charge density waves, superconducting pairing, and non-Fermi liquid behaviors.

 To construct an effective model for understanding the low-energy physics of La$_3$Ni$_2$O$_7$, a natural starting point is a bilayer two-orbital Hubbard model incorporating the Ni-$3d_{x^2-y^2}$ and $3d_{z^2}$ orbitals, derived by Wannier downfolding the DFT band structures~\cite{yaoprlmodel}. The Hamiltonian can be written as:
 $    H = H_t +H_U, $
 where $H_t$ represents the tight-binding Hamiltonian and $H_U$ represents the Coulomb interactions of Ni-$e_g$ orbitals. The $H_U$ term includes intra-orbital ($U$) and inter-orbital ($U^\prime$) Coulomb repulsion, as well as Hund's coupling ($J$). These coupling constants are related by $U^\prime = U - 2J$, reflecting Hund's rule.

\begin{figure}[t]
\includegraphics[width=\columnwidth]{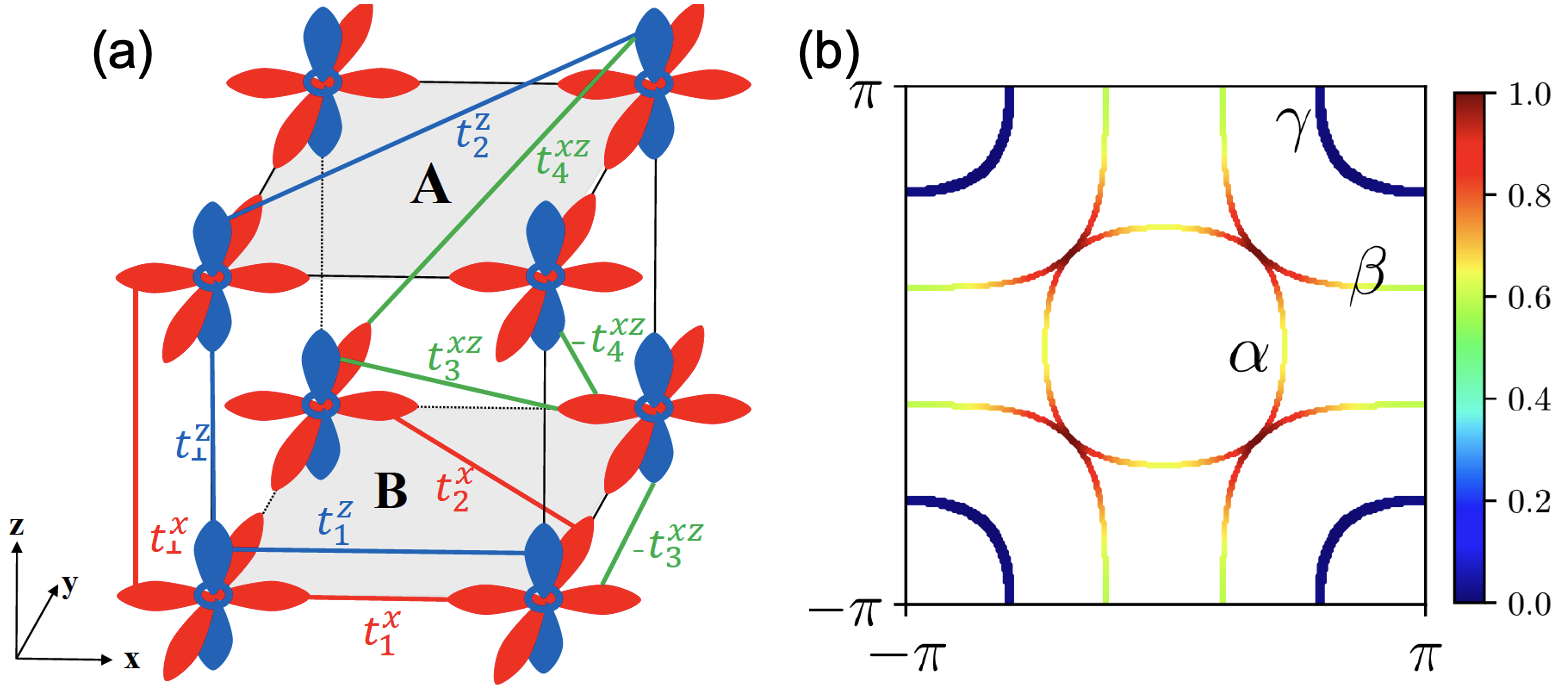}
  \caption{\label{models} (a) Hopping integrals in the bilayer two-orbital model of La$_3$Ni$_2$O$_7$ with the Ni-$d_{x^2-y^2}$ (red) and $d_{3z^2-r^2}$ (blue) orbitals. (b) Fermi surfaces determined by the bilayer two-orbital model. Reprinted from \cite{yaoprlmodel}.}
\end{figure}

 As illustrated by Fig.~\ref{models}, the tight-binding Hamiltonian $H_t$ contains the electron hopping between two $d_{z^2}$ orbitals in the adjacent layers with a large hopping constant $\sim 0.64$ eV, which opens a bonding and anti-bonding gap of $\sim 1.28$ eV. Ni $3d_{x^2-y^2}$ within each NiO layer behaves similarly as in cuprate superconductors. Furthermore, there is a hopping $\sim 0.24$ eV between the $3d_{x^2-y^2}$ and $3d_{z^2}$ orbitals within each layer. This model captures the key ingredients of La$_3$Ni$_2$O$_7$, such as the band structure and Fermi surface topology. If we further consider O-$p$ orbitals, an 11-orbital model can be constructed. Through this model, we can see significant hybridization between Ni-$3d_{z^2}$ and O-$p_z$ orbitals, as well as between the in-plane $d_{x^2-y^2}$ and O-$p_x$/$p_y$ orbitals.

 The energy dispersions of the $\alpha$ and $\beta$ bands around the Fermi level degenerate along the two-zone diagonals $k_x= \pm k_y$. This degeneracy is a characteristic feature commonly observed in cuprate superconductors, resulting from interlayer coupling with a $c$-axis hopping integral proportional to $(\cos k_x - \cos k_y)^2$~\cite{Xiang1996, Xiang1997}. However, in La$_3$Ni$_2$O$_7$, the interlayer coupling within each unit cell is very strong, which completely alters the curvature of the $\alpha$ band, transforming its Fermi surface into a pocket surrounding the $\Gamma$ point instead of the $(\pi,\pi)$ point, as seen in the Fermi surface of the $\beta$ band~\cite{RN736}. Similar band structures have also been discovered in three bilayer cuprate superconductors with strong interlayer couplings: La$_2$CaCu$_2$O$_6$, Pb$_2$Sr$_2$YCu$_3$O$_8$, and EuSr$_2$Cu$_2$NbO$_8$~\cite{RN736}. This similarity reveals a deep connection between La$_3$Ni$_2$O$_7$ and cuprate superconductors in their pairing mechanism.

  The calculation of site energies reveals that La$_3$Ni$_2$O$_7$ has smaller charge transfer energies, positioned between cuprate superconductors and infinite layer RNiO$_2$. Furthermore, this material contains two types of spin singlets: one corresponds to the Zhang-Rice singlets seen in cuprate superconductors, and the other corresponds to the antiferromagnetic correlation of interlayer $3d_{z^2}$ orbitals, with renormalized narrow bands~\cite{yaohighTc}. Doped holes occupy primarily the in-plane O-$p_x$/$p_y$ orbitals, followed by the $3d_{x^2-y^2}$ orbitals, behaving as in charge transfer gapped system. The doped holes have a significantly lower concentration in the interlayer $p_z$ and $d_{z^2}$ orbitals~\cite{Wu2024}. These results indicate the importance of oxygen orbitals in this system.

 From the DFT calculations, it was found that two Ni $3d_{z^2}$ orbitals along the c-axis are strongly coupled by the superexchange interaction, mediated by interlayer O $p_z$ orbitals. The coupling constant ranges from 0.15 to 0.2 eV. On the ab plane, cluster dynamics mean field calculations show that the antiferromagnetic superexchange interaction between $3d_{x^2-y^2}$ orbitals is about half of the interlayer one~\cite{Wu2024,yaohighTc}. 

\subsection{Pairing symmetry}

 The models proposed to account for the pairing symmetry of high-pressure bilayer La$_3$Ni$_2$O$_7$ can be classified into two groups:

 The first assumes that the $\gamma$ band emerging above the Fermi surface drives high-T$_c$ superconductivity through the interlayer Ni $3d_{z^2}$ antiferromagnetic interactions. This assumption relies on the observation that the $\gamma$-band rises above the Fermi level concurrently with the onset of the superconducting phase under high pressures. Given the volume size of the $\gamma$-band Fermi surface is small, these models generally predicted an $s$-wave pairing in this band. Consequently, the whole system would exhibit an s$\pm$-wave pairing symmetry due to interplay between the $\gamma$ and other bands~\cite{wangqh2023, gu2023effective, yaohighTc, PhysRevLett.131.236002, ShenQin2023, XueWang2024}. In a two-component theory~\cite{yang2023interlayer}, cooper pairs are first formed on the interlayer $3d_{z^2}$ orbitals, then phase coherence happens within the NiO$_2$ plane through the hybridization of $3d_{x^2-y^2}$ and $3d_{z^2}$ orbitals, leading to a macroscopic superconducting state.

 The second assumes that superconducting correlation is primarily driven by the pairing interaction in the $\alpha$ and $\beta$ bands that comprise most of the Fermi surfaces. In both $\alpha$ and $\beta$ bands, Ni $3d_{x^2-y^2}$ and O $2p$ electrons dominate the low-energy density of states around the Fermi level. In this case, a pairing state with a dominant $d$-wave character is expected, arising mainly from the superconducting pairing in the $\beta$ band~\cite{yaohighTc, Jiang__2024, liu2023role, RN736}. This expectation arises because the $\beta$-band shares a similar Fermi surface contour with optimally doped cuprate superconductors. The role of the $\gamma$ band is to adjust the hole-doping level in the $\alpha$ and $\beta$ bands. Specifically, the hole doping changes from highly overdoped at ambient pressure to a level at high pressure corresponding roughly to optimal doping in the cuprate superconductors, which is how the $\beta$ band achieves a high-T$_c$ superconducting state. Moreover, as the $\alpha$-band alone favors an $s$-wave pairing state, the interplay between these two bands could result in a ($d+is$)-pairing state in the parameter range relevant to the material~\cite{RN736}. In this group, an alternative scheme is to assume that the $\alpha$ and $\beta$ bands drive the superconductivity, but it further assumes the existence of a strong interlayer pairing interaction between Ni $3d_{x^2-y^2}$ electrons induced by the Hund's coupling~\cite{PhysRevLett.132.146002, PhysRevLett.132.036502, ouyang2023hund}. 

 DFT calculations have indicated that electron-phonon coupling alone in La$_3$Ni$_2$O$_7$ is insufficient to induce superconductivity, suggesting that the Cooper pairing mechanism is unconventional and may originate from antiferromagnetic fluctuations~\cite{zhan2024cooperation}. Nevertheless, the bilayer two-orbital model suggests that electron-phonon coupling may enhance the pairing potential induced by antiferromagnetic interactions. Further studies are needed to explore the role of electron-phonon coupling in the pairing mechanism.

 The entire family of bilayer 327-type nickelates by replacing La with other rare earth elements, $R_3$Ni$_2$O$_7$, has also been investigated theoretically~\cite{zhang2023trends}. The study found that electron correlation decreases from La to Lu, while the crystal-field splitting between Ni $3d_{x^2-y^2}$ and $3d_{z^2}$ remains nearly constant. The s$\pm$-wave pairing is predominant across all candidates if the pairing interaction contributes mainly by Ni 3$d_{z^2}$ electrons, and $T_c$ is expected to decrease as the radius of the rare-earth ions decreases. Hence, growing bilayer nickelates on substrates with larger in-plane lattice spacings could potentially increase $T_c$.
 
\section{Concluding remarks}
\label{Sec:Sum}

 Research into the physical properties of high-T$_c$ superconductor La$_3$Ni$_2$O$_7$ under high pressures has made significant progress during the past year ever since its discovery, opening new avenues for understanding high-temperature superconductivity. However, several fundamental challenges must be overcome to uncover the intriguing physics of this material.

 First, improving the sample quality of La$_3$Ni$_2$O$_7$ is not just a necessity but a crucial step toward understanding its superconducting properties, particularly the origin of pairing interactions. The presence of mixed phases and structural defects not only obscure the intrinsic properties of La$_3$Ni$_2$O$_7$ but also complicate our efforts to study and optimize its physical effects. Efforts should be made to purify the phase of La$_3$Ni$_2$O$_7$, enhancing its homogeneity in crystal structure and oxygen content. This effort is crucial for uncovering the physical origin of superconducting pairing and examining the theoretical models proposed. Furthermore, efforts should also be devoted to investigating doping strategies and chemical substitutions to purify the superconducting phase and increase its superconducting volume fraction.

Currently, the limited availability of experimental tools for reliable high-pressure measurements presents another significant challenge. Exploring the potential for achieving superconductivity in La$_3$Ni$_2$O$_7$ at lower pressures or even under ambient conditions could significantly mitigate this challenge, potentially leading to a substantial breakthrough in the search for new high-T$_c$ superconductors. Simultaneously, efforts should be directed toward developing new experimental methods applicable to high pressures.

Naturally, there is also interest in exploring other members of the Ruddlesden-Popper series and related compounds to uncover new high-temperature superconductors. Recently, Ac$_3$Ni$_2$O$_7$, La$_2$BaNi$_2$O$_6$F, and La$_2$SrNi$_2$O$_6$F were proposed to be superconducting at ambient pressure and can be used to testify theoretical models for bilayer nickelate superconductivity~\cite{wu2024ac3ni2o7}. Moreover, DFT calculations show that their $I4/mmm$ structures are energetically and dynamically stable at ambient or small applied pressures. Their electronic structure and magnetic properties are similar to high-pressure La$_3$Ni$_2$O$_7$.

\section*{Acknowledgments}
This work was supported by the National Natural Science Foundation of China (Grants No. 92165204, 12174454, 12488201, 12325403), the National Key Research and Development Program of China (Grants No. 2023YFA1406500, 2022YFA1602601, 2022YFA140280, 22022YFA1403201), the Guangdong Basic and Applied Basic Research Funds (Grants No. 2024B1515020040, 2021B1515120015), Guangzhou Basic and Applied Basic Research Funds (Grant No. 2024A04J6417), and Guangdong Provincial Key Laboratory of Magnetoelectric Physics and Devices (Grant No. 2022B1212010008). 


\bibliography{reference}
\end{document}